\documentclass[letter]{aa} % for the letters 
\usepackage{graphicx}
\usepackage{lscape}
\usepackage{natbib}
\bibpunct{(}{)}{;}{a}{}{,} % to follow the A&A style
\newcommand\msun{\rm\,M_\odot}

\usepackage{txfonts}
\begin{document}

   \title{How tidal erosion has shaped the relation between globular cluster specific frequency and galaxy luminosity}

\author{
                S. Mieske
          \inst{1}
          \and
          A. H. W. K\"upper \inst{2,3}
          \and
          M. Brockamp
          \inst{4}}

   \offprints{S. Mieske}

       \institute{
               European Southern Observatory, Alonso de Cordova 3107, 
Vitacura, Santiago, Chile
          \and
              Department of Astronomy, Columbia University, 550 West 120th Street, New York, NY 10027, USA
              \and
              Hubble Fellow
          \and
               Helmholtz-Institut f\"ur Strahlen- und Kernphysik, Universit\"at Bonn, Nussallee 14–16, D-53115 Bonn, Germany
}

   \date{}

%\abstract{}{}{}{}{} 
% 5 {} token are mandatory
 
\abstract 
%context heading (optional) 
{}
% aims heading  (mandatory) 
{We quantify to what extent tidal erosion of globular clusters (GCs) has contributed to the observed u-shaped relation between GC specific frequencies $S_N$ and host galaxy luminosity $M_V$.}
% methods heading (mandatory) 
{We used our \textsc{Muesli} code to calculate GC survival rates for
  typical early-type galaxy potentials covering a wide range of
  observed galaxy properties. We do this for isotropic and radially
  anisotropic GC velocity distributions.}
% results heading (mandatory) 
{We find that the calculated GC survival fraction, $f_{\rm s}$,
  depends linearly on the logarithm of the 3D mass density,
  $\rho_{3D}$, within the galaxy's half light radius, with $f_{\rm s}
  \propto \rho_{3D}^{-0.17}$. For a given galaxy, survival rates are
  lower for radially anisotropic configurations than for the isotropic
  GC cases. We apply these relations to a literature sample of 219
  early-type galaxies from Harris et al. (\citeyear{Harris2013}) in
  the range $M_V=[-24.5:-15.5]$ mag. The expected GC survival fraction
  ranges from 50\% for the most massive galaxies with the largest
  radii to 10\% for the most compact galaxies. We find that
  intermediate luminosity galaxies $M_V=[-20.5:-17.5]$ mag have the
  strongest expected GC erosion. Within the considered literature
  sample, the predicted GC survival fraction therefore defines a
  u-shaped relation with $M_V$, similar to the relation between
  specific frequency $S_N$ and $M_V$. As a consequence, the u-shape of
  $S_N$ vs.  $M_V$ gets erased almost entirely when correcting the
  $S_N$ values for the effect of GC erosion.}
% conclusions heading (optional),  leave it empty if necessary 
{Tidal erosion is an important contributor to the u-shaped relation
  between GC specific frequency and host galaxy luminosity. It must be
  taken into account when inferring primordial star cluster formation
  efficiencies from observations of GC systems in the nearby universe.}

\titlerunning{Flattening the u}


   \maketitle 
%
%________________________________________________________________

\section{Introduction}

\begin{figure*}[]
\begin{center}
\includegraphics[width=5.8cm]{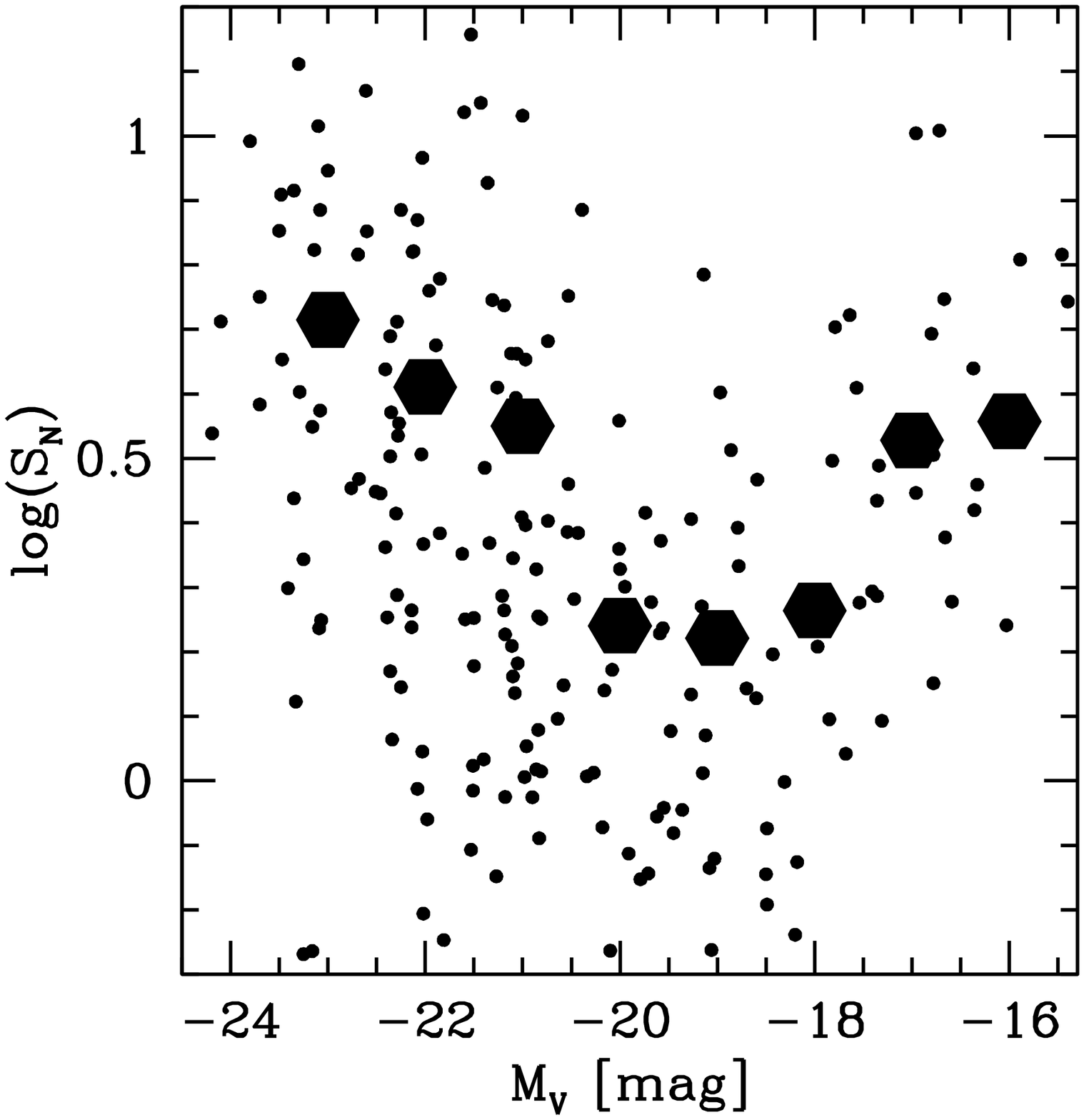}
\includegraphics[width=5.8cm]{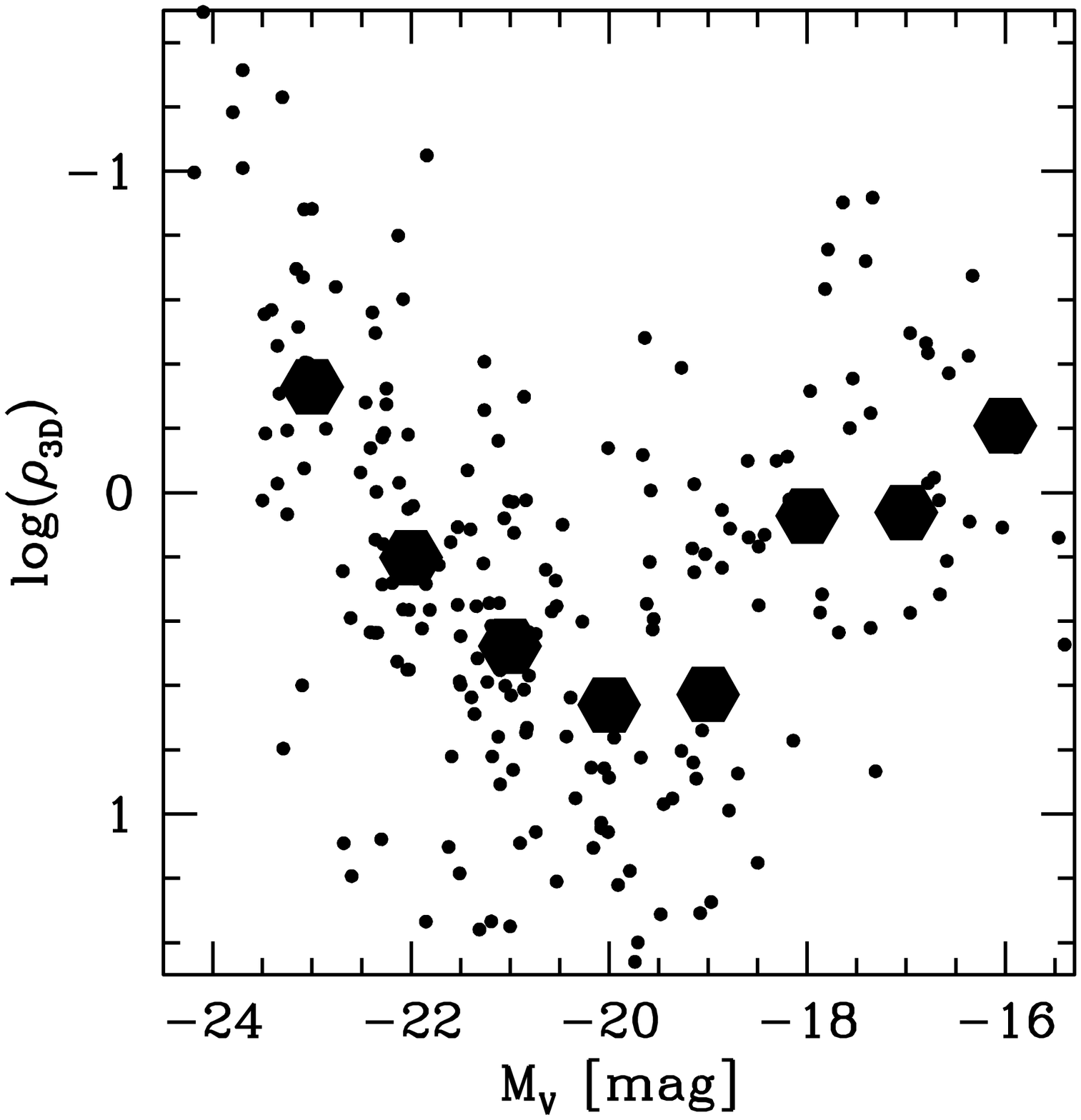}
\includegraphics[width=5.8cm]{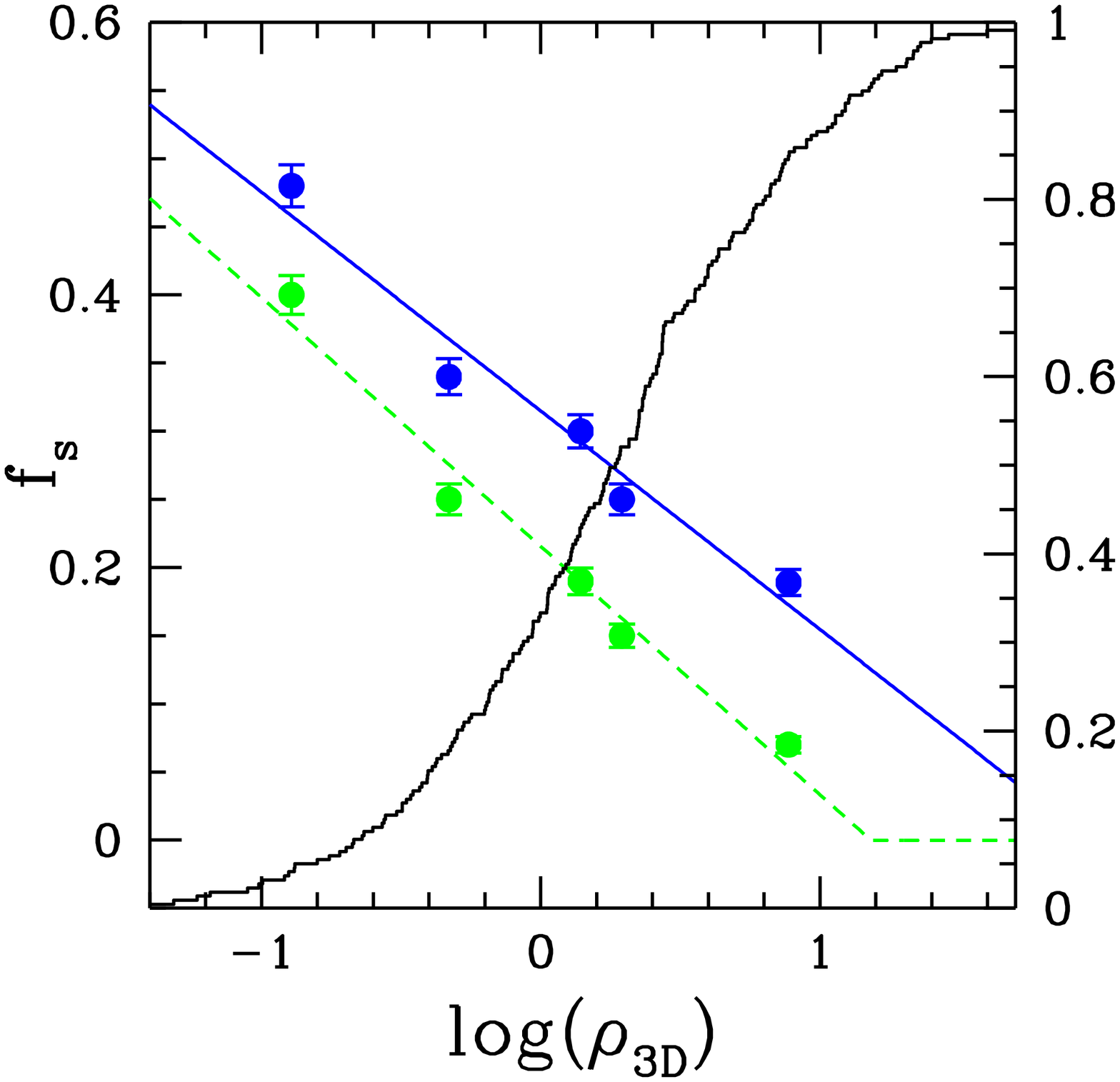}
  \caption{GC specific frequencies and galaxy mass densities define
    similar u-shape relations with $M_V$. {\bf Left}: Logarithm of the
    specific frequency, $S_N$, of GCs plotted vs.~the host galaxy
    luminosity, $M_V$. The sample is taken from the compilation of
    Harris et al. (2013), but restricted to early-type galaxies (E/SO,
    dE/SO) for which dynamical mass estimates exist. We also indicate
    the mean $S_N$ in bins of one magnitude (thick hexagons). {\bf
      Middle}: Logarithm of the projected 3D mass density within the
    half-light radius, $\rho_{3D}$, plotted vs.~$M_V$ for the same
    sample as in the left panel. The units of the mass density are
    $\msun$ pc $^{-3}$. {\bf Right:} Survival fraction, $f_S$, of GCs
    with masses $>10^5 \msun$ after 10 Gyr of evolution plotted
    vs. the logarithm of the projected 3D mass density, $\rho_{3D}$,
    within the half-light radius of their host galaxy. The data are
    based on the simulations presented in Brockamp et al. (2014). The
    blue filled circles (fit by a solid blue line)
    correspond to the assumption of an isotropic GC velocity
    distribution, whereas the green filled circles (fit by a dashed
    green line) correspond to a modestly radially biased GC
    configuration.  From the high to low survival fraction, the data
    correspond to the following template simulations: NGC 4889, VCC
    1073, IC1459, NGC4494, and NGC4564 (see also the summary in Table
    1).  The black solid curve shows the cumulative $\rho_{3D}$
      distribution of the sample of early type galaxies considered in
    the left and middle panels.}
\label{SN_force}
\end{center}
\end{figure*}

Globular clusters (GCs) are the fossil remnants of star formation in the
very early universe.  Interestingly, it has been found that the
observed specific frequency $S_N = N_{GC} \cdot 10^{0.4\,(M_V+15)}$ of
GCs in galaxies systematically changes as a
function of host galaxy luminosity $M_V$ (e.g. Forbes et
al. \citeyear{Forbes2005}, Brodie \& Strader \citeyear{Brodie2006},
Peng et al. \citeyear{Peng2008}, Spitler et al. \citeyear{Spitle2008},
Georgiev et al. \citeyear{Georgi2010}, Harris et
al. \citeyear{Harris2013}). It is high for very luminous and faint
galaxies and comparably low for intermediate luminosity galaxies. In
Fig.~\ref{SN_force} this is shown for the 219 early-type galaxies
(E/SO \& dE/SO) from the recent compilation of Harris et
al. (\citeyear{Harris2013}) for which estimates of $S_N$ and dynamical
mass exist. The luminosity range $M_V=[-24.5:-15.5]$ mag covered
  by the sample corresponds to a dynamical mass range $M_{\rm dyn}
  \sim 10^{12.5}$ to $M_{\rm dyn} \sim 10^{8.5} \msun$.

Discussion of this u-shape has focussed on systematic
variations of field star vs. star cluster formation efficiencies
across galaxies. Indeed, the observed number of GCs {\it
  per total galaxy halo mass} (i.e. including the full dark matter
halo) has a significantly smaller variation than the specific frequency normalised by galaxy luminosity (Spitler et al. \citeyear{Spitle2008},
Georgiev et al. \citeyear{Georgi2010}, Harris et
al. \citeyear{Harris2013}, Hudson et
al. \citeyear{Hudson2014}). As a result, one commonly cited explanation for
the u-shape is that the GC formation efficiency is close to uniform
per unit total halo mass, while field star formation was partially
suppressed in very massive haloes and in dwarf galaxies (Dekel \&
Birnboim \citeyear{Dekel2006}). The latter is supported by a
similar u-shape of dynamical mass-to-light ratios for galaxies (Peng et
al. \citeyear{Peng2008}).

However, both Peng et al. (\citeyear{Peng2008}) and Georgiev et
al. (\citeyear{Georgi2010}) stress that the specific frequencies of GCs
observed in the nearby universe are a combination of (at least) two
effects: the ratio of star-cluster and field-star formation
efficiencies, and the GC destruction rate. Indeed, Murali \&
  Weinberg (\citeyear{Murali1997}) and Vesperini
  (\citeyear{Vesper2000}) have already suggested that the tidal destruction of
  GCs is an important contributor to observed variations
  in specific frequency with host-galaxy properties. Several more
  studies have since investigated the destruction of GCs in an
  external tidal field (e.g. Baumgardt \citeyear{Baumga1998}, Fall \&
  Zhang \citeyear{Fall2001}, Vesperini et al. \citeyear{Vesper2003},
  Sanchez-Janssen et al. \citeyear{Sanche2012}, Smith et
  al. \citeyear{Smith2013}). However, a quantitative evaluation of GC
  erosion as a function of galaxy properties has been difficult up to
  now because of the numerical challenges associated with a full treatment
  of the problem. There are several internal (stellar evolution,
two-body relaxation) and external (tidal forces, dynamical friction)
processes acting simultaneously on GCs, with
efficiencies depending strongly on the respective clusters' masses and
orbits.

In Brockamp et al. (\citeyear{Brocka2014}), the tidal erosion of GC
systems in elliptical galaxies is systematically investigated with
$N$-body simulations for a wide range of host galaxy masses and radii,
as well as different initial GC velocity distributions.  One of the
main findings of this study is that, after 10 Gyr of dynamical
evolution, very compact galaxies like M\,32 have GC survival fractions
of only a few percent, while very extended giant elliptical galaxies
have survival fractions of up to 50\%. Futhermore, it is found that GC
number density profiles are centrally flattened in less than a Hubble
time, naturally explaining the observed cored GC distributions that
are generally more extended than the galaxy star light (e.g. Peng et
al.~\citeyear{Peng2011}, Forbes et al.~\citeyear{Forbes2012}).

The aim of the present letter is to quantify the importance of
  the tidal erosion of GC systems in shaping the observed relation
between the specific frequency of globular clusters, $S_N$, and a
galaxy's absolute luminosity, $M_V$. Based on the results of Brockamp
et al. (\citeyear{Brocka2014}), we compare the predicted survival
fraction of GCs in early-type galaxies to the observed specific
frequency of GCs using the recent compilation of Harris et
al. (\citeyear{Harris2013}) and discuss our findings.

\section{Erosion of GC systems}

In the middle panel of Fig.~\ref{SN_force} we plot the logarithm of
the 3D dynamical mass density within the half light radius,
$\rho_{3D}$, versus the absolute luminosity, $M_V$, for the same
compilation from Harris et al. (\citeyear{Harris2013}) as in the
left-hand panel. A u-shape trend similar to the left-hand panel is
seen. As shown in the following, the GC survival fraction depends
linearly on $\log(\rho_{3D})$, implying that part of the u-shape in
$S_N$ vs. $M_V$ arises from GC erosion.

\subsection{A simple formula derived from Brockamp et al. (2014)}

In Brockamp et al. (2014), we presented the versatile \textsc{Muesli}
code for investigating the erosion of GC systems in
elliptical galaxies.  \textsc{Muesli} follows the orbits of individual
clusters in galaxy potentials and applies both internal and external
dissolution processes to them, yielding GC destruction rates as a
function of host galaxy parameters. In a first application, we studied
spherical galaxies with different masses, sizes, central super-massive
black holes, density profiles, and velocity distributions. We
found that the erosion efficiency strongly depends on the compactness
of the galaxy and the anisotropy of its velocity distribution.

For the present study, we consider S\'{e}rsic galaxy models that have
concentration parameters n=4 (and n=2.99 for VCC\,1073) with both
initially isotropic and initially radial-biased velocity distributions
for the GC orbits. The anisotropic velocity distribution follows
Osipkov-Merritt models (Osipkov~\citeyear{Osipko1979},
Merritt~\citeyear{Merrit1985}) with a characteristic anisotropy radius
that equals the 3D half mass radius ($\approx 1.35$ times the
projected half light radius, $r_h$). This choice of parameter yields a
moderately anisotropic velocity distribution, with the central parts
still isotropic and the outer parts more radially biased (see
e.g.~figure A1 in Brockamp et al.~\citeyear{Brocka2014}). At
the end of the simulations after 10 Gyrs, the central orbital
distribution has become tangentially biased due to the higher
destruction of GCs on radial orbits.

We assume that the galaxies have no central SMBH, because we found in
Brockamp et al. (2014) that these have very little effect on GC
erosion in general. The total number of GCs at the start of the
simulations was 20000, sampled from a power-law mass function reaching
down to $10^4 \msun$ (see Brockamp et al.~for details). To establish
comparability between our simulations and the typical completeness
limit for observations of GC systems (e.g. Harris et al. 2013), we
restricted ourselves to simulated clusters with masses above $10^5
\msun$. To cover the full range of 3D mass densities from Harris et
al. (2013), we made two new simulations (representing NGC 4564 and VCC
1073) following the setup described in Brockamp et al.~(2014). A
summary of the simulation results for the five galaxies can be found
in Table~\ref{table_erosion}.

The results can be expressed as a relation between GC
survival fraction and the logarithm of the 3D dynamical mass density
within the half light radius. In Fig.~\ref{SN_force} (right
  panel), we plot the GC survival fraction, $f_s$, versus the
logarithm of the 3D mass density within the half-light radius,
$\rho_{3D} \equiv \mbox{M}_{dyn}/(r_h^3)$. It is
apparent that radial anisotropy leads to a higher GC destruction rate
as more clusters orbit through the denser central parts of the galaxy.
In Fig.~\ref{SN_force} we indicate a linear fit to the five data
points for both cases of isotropy and radial anisotropy. These linear
fits are given by

\begin{equation}
f_{\rm s,iso}=-0.160\cdot\log_{10}(\rho_{3D})+0.315
\label{eq:force_iso}
\end{equation}
\begin{equation}
f_{\rm s,aniso}=-0.182\cdot\log_{10}(\rho_{3D})+0.216   .
\label{eq:force_aniso}
\end{equation}

\begin{figure*}[]
\begin{center}
\includegraphics[width=8.6cm]{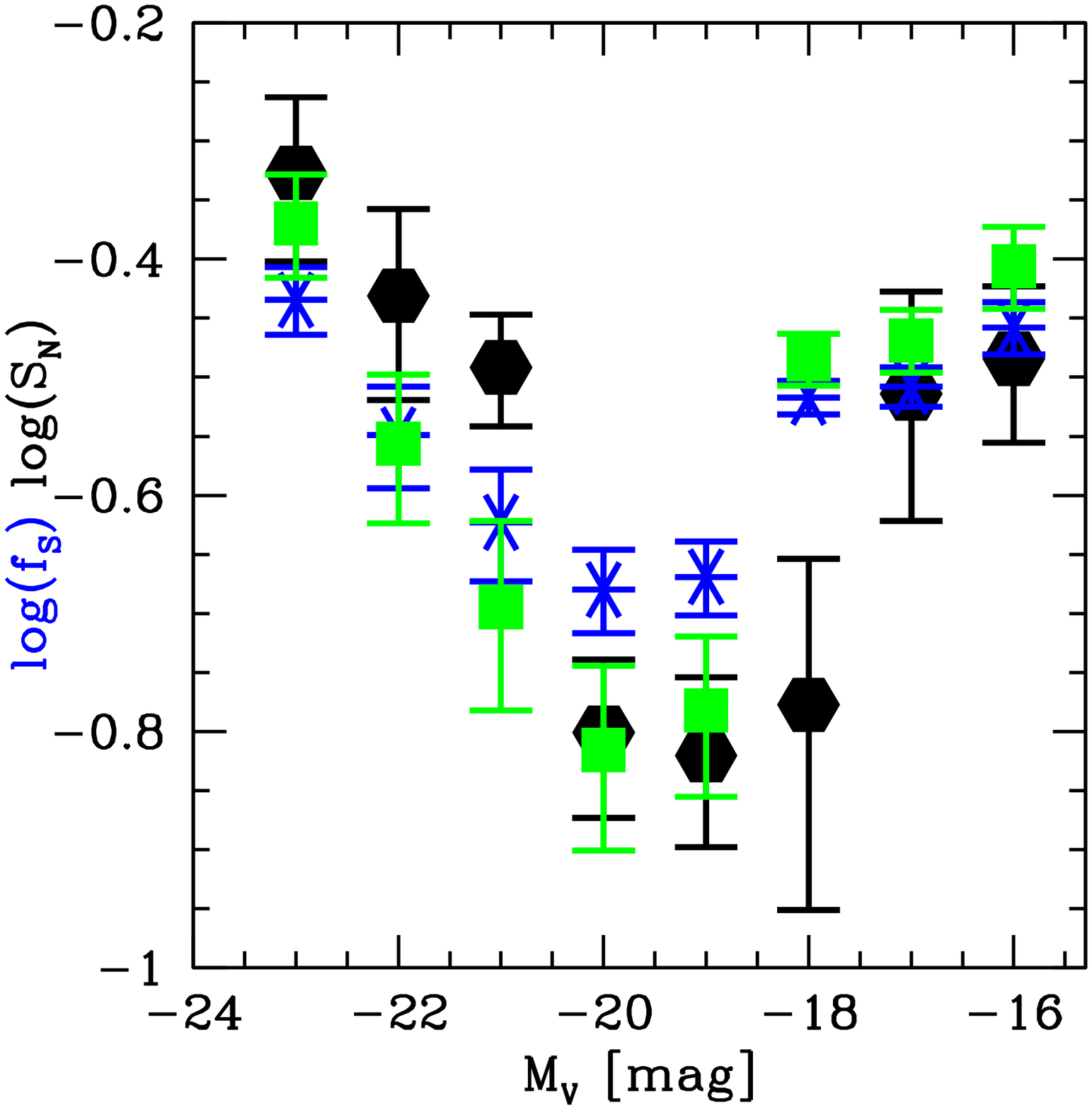}
\includegraphics[width=8.6cm]{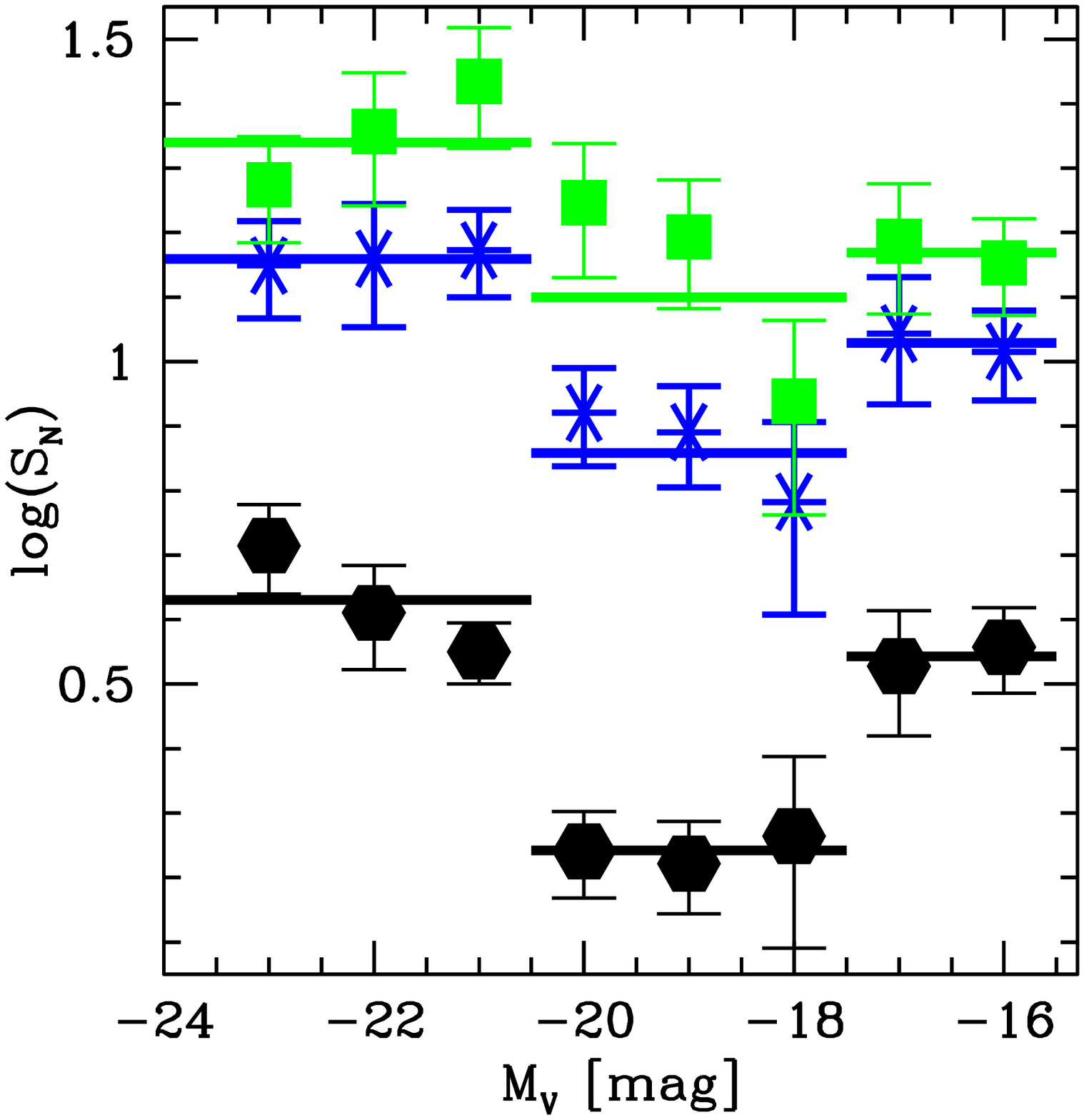}

\caption{How tidal erosion changes the relation between GC specific
  frequency $S_N$ and galaxy luminosity $M_V$. {\bf Left}: We compare the
  observed GC specific frequencies, $S_N$ (offset by a constant
  factor), with the predicted GC survival fractions, $f_S$, for the
  same set of galaxies from Fig.~\ref{SN_force}, using the predictions
  from Eqs.~\ref{eq:force_iso} and ~\ref{eq:force_aniso}. Large
  symbols indicate means within 1 mag bins. The observed $S_N$ are
  indicated in (black) hexagons, survival fractions in (blue)
  asterisks for isotropic orbits and in (green) squares for radially
  anisotropic orbits. {\bf Right:} Predicted primordial specific
  frequencies are obtained by dividing the observed specific
  frequencies by the survival fractions of the left panel. Colours and
  symbols as in the left panel. Black hexagons are again the
    observed present-day specific frequencies. The u-shape is
  attenuated for the isotropic case and is almost erased for the
  anisotropic case, see text for details. Horizontal thick lines denote the average $S_N$ values within the three considered luminosity ranges $M_V=[-24.5:-20.5]$ (high luminosity), $M_V=[-20.5:-17.5]$
(intermediate luminosity), $M_V=[-17.5:-15.5]$ (low luminosity). }
\label{survive}
\end{center}
\end{figure*}

We note that we do not include our simulation for M\,32 at very high
density, since it is in a density regime where the GC survival
fraction is already `saturated' at 0. The histogram in
Fig.~\ref{SN_force} shows the cumulative distribution of 3D mass
densities, $\rho_{3D}$, for the literature sample that defines the
$S_N$ vs. $M_V$ relation in Figs.~\ref{SN_force} and~\ref{survive}. The
range of 3D density of the simulated GC systems covers the range of
the literature sample.

\begin{table}
\caption{\label{table_erosion}\small{Main parameters of the five
    GC-system erosion simulations shown in Fig.~\ref{SN_force}. Half light radius, $r_h$, is given in kpc, dynamical
    galaxy mass, M$_{\rm dyn}$, is given in units of 10$^9 \msun$. The
    survival fraction, $f_s$, adopted here is the ratio of the number of
    GCs with mass M$\,>10^5 \msun$ after 10 Gyr to the number of GCs
    with mass M$\,>10^5 \msun$ at the start of the simulation. We show
    the results both for the isotropic and the radially anisotropic
    case. }}
\begin{center}
\begin{tabular}{l|l|l|l|l|l|l}
Galaxy  & $r_{h}$& M$_{\rm dyn}$ & $\log(\rho_{3D})$ & $f_{\rm s,iso}$ & $f_{\rm s,aniso}$ &$M_V$ \\\hline
NGC\,4564 &       1.78         &           43.5            &   0.89    &        0.19    &    0.07 & -20.0  \\
NGC\,4494	&	3.72	&		100		& 0.29	&	0.25 &	0.15	&-22.0 \\
IC\,1459	&	6.0	&		300		&0.14	&	0.30 &	0.19	&-22.7\\
VCC\,1073	&	2.0		&	3.75		&-0.33	&	0.34	&0.25	&-17.5 \\
NGC\,4889	&	25.0	&		2000		&-0.89	&	0.48 &	0.40	&-24.2\\
\end{tabular}
\end{center}
\end{table}

\subsection{Application to the sample of Harris et al.}

In the left-hand panel of Fig.~\ref{survive}, we have calculated the
GC survival fraction for the early-type galaxies in the sample of
Fig.~\ref{SN_force}, using the observed galaxy mass densities and the
relation between survival fraction and mass density from
equations~\ref{eq:force_iso} and ~\ref{eq:force_aniso}. For the sake
of clarity we focus on average values within 1 magnitude bins and omit
individual data points. We indicate the logarithm of the expected
survival fraction for an isotropic velocity distribution
(equation~\ref{eq:force_iso}) and a radially biased velocity
distribution (equation~\ref{eq:force_aniso}). For comparison we
over-plot the present-day specific frequencies (same data points as
in the left plot of Fig.~\ref{SN_force}), shifted by a constant offset in log
space.

The most striking feature of Fig.~\ref{survive} is that the
  expected survival fraction defines a u-shape that is very similar to
  the observed $S_N$ vs. $M_V$ data points. For the case of
isotropic GC velocity distributions, the amplitude of variation in the predicted
survival fractions is about a factor of two lower than the observed
variation of specific frequencies: the ratio between the
maximum and minimum survival rates within the eight magnitude bins is
1.8. For comparison, the ratio between maximum and minimum specific
frequency $S_N$ within the same set of eight magnitude bins is 3.1
(5.2 to 1.7). Interestingly, for the case of radial anisotropy, 
the amplitude of the predicted u-shape relation matches the observations well. 
For this case, the ratio between the maximum
and minimum survival rate within the eight magnitude bins is 2.8,
matching the observed amplitude variation to within 10\%.

%M_V=[-20.5:-17.5]$ is about 2.3 times lower [1.7 vs. 4.0] than
%the mean $S_N$ for both brighter $M_V=[-24.5:-20.5]$ and fainter
%galaxies $M_V=[-17.5:-15.5]$.

In the right-hand panel we show the `primordial' specific frequencies of
the same galaxies, obtained by dividing the observed $S_N$ values
by the survival fractions. We consider the three subsamples
$M_V=[-24.5:-20.5]$ (high luminosity), $M_V=[-20.5:-17.5]$
(intermediate luminosity), $M_V=[-17.5:-15.5]$ (low luminosity).  For
the case of an isotropic GC velocity distribution, the predicted
primordial specific frequencies are 14.4, 7.2, 10.7 in those three
ranges. For radial anisotropy, these predictions are 22.9, 14.0, 14.8.

\begin{table}
\caption{\label{KS}\small{KS-test probabilities that the $S_N$ distribution within the three subsamples (high, intermediate [=int.], low luminosity) have the same parent distribution. See also the right panel of Fig.~\ref{survive}. We show these probabilities for the observed $S_N$, and for the primordial $S_N$ before erosion, both for isotropic and radially anisotropic GC velocity distributions.}}
\begin{center}
\begin{tabular}{l|lll}
KS probabilities &high vs. int & high vs. low & int. vs low\\\hline
Present-day & $5\cdot10^{-6}$&0.70 &$1.5\cdot10^{-3}$\\
Prim. iso&$1\cdot10^{-4}$  & 0.30 & 0.15\\
Prim. aniso& 0.10 & 0.53 &0.44\\
\end{tabular}
\end{center}
\end{table}

In Table~\ref{KS} we show the results of a Kolmogorov-Smirnov (KS)
test comparing the $S_N$ distribution of the high, intermediate, and
low-luminosity samples. We consider the following three cases:
observed $S_N$, inferred primordial $S_N$ for both isotropic and
anisotropic GC velocity distributions.  The characteristic u-shape in
the observed $S_N$ vs. $M_V$ plane is clearly reflected in the KS
probabilities: the intermediate-luminosity sample is grossly
inconsistent with both high- and low-luminosity galaxies, while the
other two agree well. The situation gradually changes for the
primordial $S_N$ with isotropic GC orbits, and the disagreement
between the subsamples disappears for primordial $S_N$ based on
anisotropic GC orbits. The comparison between high- and
intermediate-luminosity samples for the latter case gives a comparably
high (10\%) probability that both have the same parent distribution.

\section{Summary and conclusions}

Applying the simulation results of GC erosion in elliptical galaxies
from Brockamp et al. (2014), we calculated the survival rates of
GCs in early type galaxies, using a sample of 
219 early type galaxies in the $M_V$=[-24.5:-15.5] mag. 

We show that the simulations of GC system erosion in Brockamp et
al. (2014) are approximated well by a linear relation between GC
survival fraction and the logarithm of the 3D dynamical mass density
within the galaxy's half light radius ($f_{\rm s} \propto
\rho_{3D}^{-0.17}$.).  For a given galaxy, survival rates are lower
for radially anisotropic GC velocity distributions than for isotropic
ones. Within the considered literature sample from Harris et
al. (2013), the expected GC survival fraction ranges from 50\% for
the most massive galaxies with the largest radii to 10\% for the most
compact galaxies.

Within the considered literature sample, the predicted GC survival
fraction defines a similar u-shaped relation with $M_V$ than the
specific frequency $S_N$. In particular, the
peak-to-valley of the survival fraction when considering moderate
radial anisotropy agrees to within 10\% with the peak-to-valley of the
$S_N$.  Higher degrees of primordial anisotropy may reconcile the
results completely.

 As a consequence, the u-shape of $S_N$ vs. $M_V$ becomes almost flat
 when correcting the $S_N$ values for the effect of GC erosion for
 radially anisotropic orbits. A KS test shows no significant
 difference in corrected $S_N$ distribution between high-, intermediate-
 and low-luminosity galaxies. The original u-shape is notably
 attenuated but still preserved for the assumption of an isotropic GC
 velocity distribution.

We therefore find that GC erosion is a major contributor to the
u-shaped relation between present-day GC specific frequency, $S_N$,
and host galaxy luminosity, $M_V$. It must be taken into account when
inferring primordial star cluster formation efficiencies from
observations of GC systems in the nearby universe, and may thus change
our picture of how field star and star cluster formation were coupled
in the early universe.

\vspace{0.1cm}\noindent We conclude with a note of caution. It is
unlikely that tidal erosion alone is responsible for the entire
variation in $S_N$ vs. $M_V$ we observe today. For example, our
simulations (Brockamp et al. 2014) do not explicitly include the
hierarchical nature of structure formation in the
universe. Current GC systems of most galaxies are probably
compounds of smaller GC systems acquired through mergers (e.g.~van
Dokkum et al.~\citeyear{vandok2008}). Blue and red GCs
have different spatial distributions (e.g. Brodie \& Strader
~\citeyear{Brodie2006}), an effect not yet considered in our
simulations. Furthermore, host-galaxy potentials should be evolving
with time. A detailed inclusion of such effects into simulations of GC
system dynamical evolution is thus a promising prospect.

\begin{acknowledgements} 
The authors would like to thank the anonymous referee for valuable
comments, and J.P. Ostriker for interesting discussions. We thank
Michael Marks for interesting discussions and Holger Baumgardt for
technical support. SM acknowledges support from the ESO Director
General Discretionary Fund and the kind hospitality of the Columbia
University Astronomy Department. AHWK acknowledges support through
Hubble Fellowship grant HST-HF-51323.01-A awarded by the Space
Telescope Science Institute, which is operated by the Association of
Universities for Research in Astronomy, Inc., for NASA, under contract
NAS 5-26555. MB acknowledges support through DFG grant KR 1635/39-1.
\end{acknowledgements}

\bibliographystyle{aa}
\bibliography{mieske_language}

\end{document}